\documentclass{article}

\PassOptionsToPackage{numbers}{natbib}

\usepackage{amsmath}
\usepackage{graphicx}
\usepackage{caption}
\usepackage{subcaption}
\graphicspath{ {./figures/} }

\usepackage[final]{neurips_2023}

\usepackage[utf8]{inputenc} 
\usepackage[T1]{fontenc}    
\usepackage{hyperref}       
\usepackage{url}            
\usepackage{booktabs}       
\usepackage{amsfonts}       
\usepackage{nicefrac}       
\usepackage{microtype}      
\usepackage{xcolor}         

\title{States as goal-directed concepts: an epistemic approach to state-representation learning}

\author{%
Nadav Amir$^{1*}$ \quad Yael Niv$^{1}$ \quad Angela Langdon$^2$\\
$^1$Princeton University \quad $^2$National Institute of Mental Health\\
\texttt{\{nadav.amir,yael\}@princeton.edu}\\
\texttt{angela.langdon@nih.gov}
}

\begin{document}

\maketitle

\begin{abstract}
Our goals fundamentally shape how we experience the world. For example, when we are hungry, we tend to view objects in our environment according to whether or not they are edible (or tasty). Alternatively, when we are cold, we may view the very same objects according to their ability to produce heat. Computational theories of learning in cognitive systems, such as reinforcement learning, use the notion of ``state-representation" to describe how agents decide which features of their environment are behaviorally-relevant and which can be ignored. However, these approaches typically assume ``ground-truth" state representations that are known by the agent, and reward functions that need to be learned. Here we suggest an alternative approach in which state-representations are not assumed veridical, or even pre-defined, but rather emerge from the agent's goals through interaction with its environment. We illustrate this novel perspective by inferring the goals driving rat behavior in an odor-guided choice task and discuss its implications for developing, from first principles, an information-theoretic account of goal-directed state representation learning and behavior.
\end{abstract}

\section{Introduction}
Concepts are the building blocks of mental representations, providing scaffolding for generalizations over individual objects or events. How do animals (including humans) form concepts and why do they form the particular ones that they do? These questions have a long history in both Eastern and Western philosophy \cite{hume1896treatise,wittgenstein1953philosophical,rosch1975family,siderits2011apoha}.
More recently, computational learning theories have started addressing a related question, namely, how do cognitive agents generate internal models of their environment, called ``state-representations'', generalizing over their experiences for efficient learning \cite{niv2019learning,langdon2019uncovering}? 
Here, we suggest that state-representations can be understood as concepts, formed by an agent in order to achieve particular goals \cite{dunne2004foundations}. Under this account, a concept, such as ``fire'', is formed because some set of interests, such as a desire for warmth, leads agents to construe particular entities as similar to each other in virtue of their efficacy in obtaining the goal of warming up. Drawing this parallel between ``concepts'' and ``states'', we propose that state-representations should be understood in terms of the goals they subserve. To explore this hypothesis, we develop a formal framework for describing goal-dependent state-representations and illustrate its application by inferring animals' goals from empirically observed behavior in a well-studied odor-guided choice task.

\section{Formal setting}
We assume the setting of an observation-action cycle, i.e., an agent receiving an observation from its environment and subsequently performing an action, then receiving a new observation and so on. We denote by $\mathcal{O}$ and $\mathcal{A}$ the set of possible observations and actions, respectively. An experience sequence, or \emph{experience} for short, is a finite sequence of observation-action pairs: $h=o_1,a_1,o_2,a_2,...,o_n,a_n$. For every non-negative integer, $n\geq0$, we denote by $\mathcal{H}_n\equiv(\mathcal{O}\times\mathcal{A})^n$ the set of all experiences of length $n$. The collection of all finite experiences is denoted by $\mathcal{H}=\cup_{n=1}^{\infty}\mathcal{H}_n$. In non-deterministic settings, it will be useful to consider distributions over experiences rather than individual experiences themselves and we denote the set of all probability distributions over finite experiences by $\Delta(\mathcal{H})$. 
Following \citet{bowling2022settling}, we define a \emph{goal} as a binary preference relation over experience distributions. For any pair of experience distributions, $A,B\in\Delta(\mathcal{H})$, we write $A\succeq_gB$ to indicate that experience distribution $A$ is weakly preferred by the agent over $B$ (i.e., that $A$ is at least as desirable as $B$) with respect to goal $g$. 
When $A\succeq_gB$ and $B\succeq_gA$ both hold, $A$ and $B$ are equally preferred with respect to $g$, denoted as $A\sim_gB$. We observe that $\sim_g$ 
is an equivalence relation, i.e., it satisfies the following properties:
\begin{itemize}
    \item Reflexivity: $A\sim_gA$ for all $A\in\Delta(\mathcal{H}).$
    \item Symmetry: $A\sim_gB$ implies $B\sim_gA$ for all $A,B\in\Delta(\mathcal{H}).$
    \item Transitivity: if $A\sim_gB$ and $B\sim_g C$ then $A\sim_g C$ for all $A,B,C\in\Delta(\mathcal{H}).$
\end{itemize}
Therefore, every goal induces a partition of $\Delta(\mathcal{H})$ into disjoint sets of equally desirable experience distributions. For goal $g$, we define the goal-directed, or \emph{telic}, state representation, $\mathcal{S}_g$, as the partition of experience distributions into equivalence classes it induces:
\begin{equation}
    \mathcal{S}_g= \Delta(\mathcal{H})/\sim_g.
\end{equation}
In other words, each telic state represents a generalization over all equally desirable experience distributions. This definition captures the intuition that agents need not distinguish between experiences that are equivalent with respect to their goal. Furthermore, since different telic states are, by definition, non-equivalent with respect to $\succeq_g$, the goal also determines whether a transition between any two telic states brings the agent in closer alignment to, or further away from its goal.

\section{Results: goal inference in an odor-guided choice task}
\begin{figure}[htb]
  \centering
  \includegraphics[scale=0.4]{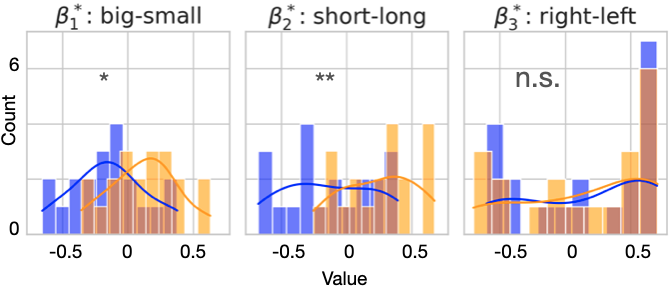}
  \caption{\textbf{Optimized weight parameters for liquid amount (left), delay duration (center) and side choice (right) preferences.} Optimized $\beta$ values maximizing the goal-alignment coefficient of empirical histories (orange) are significantly larger than those of simulated random actions yoked to the observation histories of each animal (purple) for big vs. small amount and short vs. long delay but not for right vs. left nose pokes. Solid lines show Gaussian kernel density distribution estimates. Asterisks indicate significance levels (paired t-test, $^*p<0.01$; $^{**}p<0.001$).}
  \label{fig:optimized_betas}
  \vspace{0.5cm}
  \nextfloat
  \includegraphics[scale=0.4]{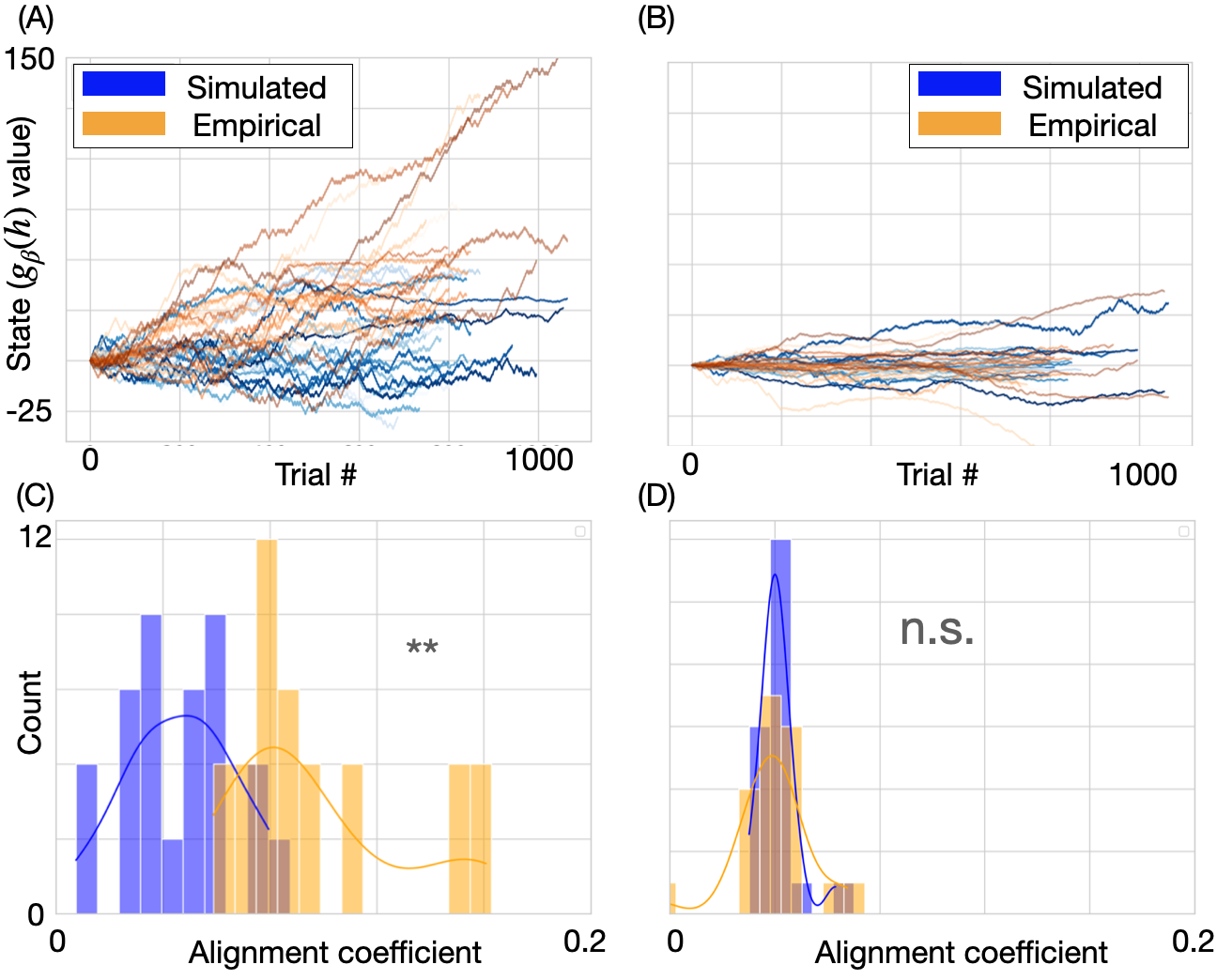}
        \caption{\textbf{State trajectories (top) and Goal Alignment Coefficients (bottom) for optimized weights (left) and random weights (right).} Top: State trajectories for optimized (A) and random (B) $\beta$ weights show the telic state, defined as the cumulative weighted sum $g_\beta$ for individual animals (ordinate) for different lengths of trial histories (abscissa). State trajectories for real animals (orange) reach larger $g_{\beta}(h)$ values than those of random simulated animals (blue). However, this is only true when using $\beta^*$ (A), not with random values of $\beta$ (B).Bottom: histograms of corresponding Goal Alignment Coefficients (GACs) of all animals (real in orange, simulated in blue) for optimized (left) and random (right) $\beta$ weights. Optimized  GACs for empirical trajectories were significantly larger than for simulated ones (paired t-test,  $p<10^{-4}$), but this is true only for optimized weights (C), not for random weights (D).}
        \label{fig:state_trajectories}
\end{figure}

We illustrate our proposed framework by inferring goal alignment from empirical behavior in a well-studied odor-guided choice task \cite{roesch2006encoding}.
Briefly, rats were trained to sample an odor at a central odor port, before responding by nose-poking in one of two fluid wells. The odor stimulus provided a cue for which of two wells would be associated with delivery of a certain amount of sucrose liquid. Two of the odors signalled “forced choice” trials, one indicating that the liquid will be available in the left well, and one indicating the right well.  A third odor—“free choice”—indicated liquid availability in either well. After liquid delivery, or choice of a non-indicated well, the rat waited for a cue indicating the start of the next trial. Importantly, if a “valid” well was chosen on any trial (i.e., the indicated well on forced-choice trials, or either well on free-choice trials), the delay to and amount of liquid was determined by the side of the well, not the odor. Unsignaled to the animal, in each block of the task, one well delivered either at a shorter delay or a larger amount than the other well; amount and delay contingencies changed between blocks during a session.
We denote the set of possible observations and actions in the task as follows:
\begin{align*}
\mathcal{O} =\{&\textit{Left Odor, Right Odor, Free Odor,} \\ 
& \textit{Long Delay, Short Delay, Small Amount, Big Amount}\},\\
\mathcal{A}=\{&\textit{Right Poke, Left Poke, Wait for Cue}\}.
\end{align*}
We define an experience for this task as a sequence of trials, each consisting of an observation-action pair, for example: 
\begin{equation*}
h=LO,LP,LD,WC,FO,RP,SD,WC,... 
\end{equation*}
where elements of $\mathcal{O}$ and $\mathcal{A}$ are denoted by their initials.  
For simplicity, we consider a simple parameterized family of goals, defined as the weighted sum of the differences between the number of right vs. left nose pokes, short vs. long delays, and big vs. small amounts, appearing in an individual experience:
\begin{equation}
\label{eq:odor_task_goal}
g_\beta(h) = \beta_1(N_{BA}^h-N_{SA}^h)+\beta_2(N_{SD}^h-N_{LD}^h)+\beta_3(N_{RP}^h-N_{LP}^h),
\end{equation}
where $N_{BA}^h,N_{SA}^h,N_{SD}^h$ and $N_{LD}^h$ denote the number of of $\textit{Big Amount, Small Amount, Short Delay}$ and $\textit{Long Delay}$ observations, respectively, and $N_{RP}^h$ and $N_{LP}^h$ denote the number of $\textit{Right Poke}$ and $\textit{Left Poke}$ actions, respectively, in experience $h$. The parameters $\beta_1\beta_2,$ and $\beta_3$, determine the relative weight of the corresponding difference between action or observation counts in determining the animal's goal. While our formalism is general with respect to the form of the goal, this particular goal family allows us to monitor preference for earlier rewards, larger rewards, and side biases -- all characteristics of animal behavior in this task.
For an empirical experience $h=o_1,a_1,...,o_n,a_n$ consisting of $n$ trials, we define the corresponding \emph{state-trajectory} as the sequence of telic states $\{S^\beta_t\}_{t=1}^n$  where $S^\beta_t$ consists of all $t$-trial histories that are $g_\beta$-equivalent to the first $t$ trials of $h$:
\begin{equation}
    S^\beta_t=\{h'\in\mathcal{H}_t : g_\beta(h')=g_\beta(o_1,a_1,...,o_t,a_t)\}.
\end{equation}
Given a goal, $g_\beta$, and an empirical experience, $h$, we quantify the alignment between $h$ and $g_\beta$ using the following Goal Alignment Coefficient (GAC):
\begin{equation}
\label{eq:GAC}
    GAC_\beta(h)=g_\beta(h)/n.
\end{equation}
For a given experience, $h$, the GAC measures the average increase in the goal value (Eq.\ref{eq:odor_task_goal}) for each trial in $h$. We used the GAC, to estimate the parameters $\beta^*$ that maximize the alignment between a given empirical set of histories $\{h_j\}_{j=1}^N$ and the goal $g_{\beta^*}$, subject to a regularization constraint on $\beta$: 
\begin{equation}
    \beta^*=\arg\max_\beta \sum_{j=1}^N GAC_\beta(h_j)\text{ s.t } \lVert \mathbb{\beta} \rVert^2=1,
\end{equation}
 where we impose the regularization constraint $\lVert \beta \rVert^2 =\sum_{i=1}^3\beta_i^2=1$ on the weight parameters to fix the scale of $g_\beta$. In other words, given the empirical experience histories of an individual animal, and a class of $\beta$-parameterized goals, we estimated the parameter values, $\beta^*$ such that the animal's behavior is maximally aligned with the goal $g_{\beta^*}$. The $\beta^*$ values for all animals are plotted in Fig.\ref{fig:optimized_betas} (orange histograms). As a baseline for comparison, we also plotted $\beta^*$ values for simulated animals using the same odor observations and liquid-outcome contingencies as the empirical data but with randomly selected left or right action choices (blue histograms). This analysis revealed that goals fit to the empirical behavior had significantly larger weights on the difference between the number of short vs. long delays and big vs. small liquid amounts compared to goals fitted to simulated random behavior. 
 Next, we plotted the value of $g_\beta$ for the state trajectories defined by cumulative trial sequences of the empirical and simulated histories (Fig.\ref{fig:state_trajectories} top, orange and blue lines respectively). We compared the trajectories obtained using the optimized weights, $\beta^*$, (left) with those obtained using random weights sampled from a unit normal distribution, $\beta\sim\mathcal{N}(0,1)$, (right). We also plotted the GACs  for the empirical and simulated animals (Fig.\ref{fig:state_trajectories} bottom, orange and blue histograms), comparing optimized (left) and random (right) weights. Weight optimization significantly increased the goal-alignment for the empirical but not for the simulated animals, demonstrating the sensitivity of the GAC as as general-purpose measure of goal-directed behavior.  

\section{Discussion}
\subsection{Relation to previous work}
The need for a formal theory of learning in cognitive systems that centers on agent's goals has recently been called into attention \cite{molinaro2023goal}. 
The current work provides a step in this direction, based on the definition of telic states as equivalence classes over experience sequences with respect to goals. While the notion of states as equivalence classes is not new \cite{minsky1967computation}, we suggest that goals, defined as preference relations over experience sequences \cite{bowling2022settling}, provide an natural epistemic foundation for constructing state representations in cognitive systems by generalizing over goal-equivalent experiences. Our approach differs from previous theoretical accounts of goal-directed state abstraction in important ways. First, while previous accounts typically assume a pre-given ``ground" state representation and reward function and describe how various generalizations across these facilitate planning and decision-making in certain contexts (see \cite{li2006towards} for a review), our approach does not assume any a-priori state representation or reward structure but rather posits that state representations are fully goal dependent. Thus, whereas previous goal-conditioned reinforcement-learning models view goals as a subset of future states that the agent wants to reach \cite{kaelbling1993learning}, our framework goes in the opposite direction by starting with the goal and deriving the state-representation that best alligns with it. 
In some sense, our approach ``shifts the burden" from explaining state representations to explaining sensory-motor ones, as it assumes a pre-defined set of possible actions and observations available to the agent. However, we argue that sensory-motor representations themselves may be similarly explained in terms of underlying goals, as evinced by recent studies on ``motivated perception" \cite{balcetis2006see,leong2019neurocomputational} and goal-conditioned action representations \cite{fogassi2005parietal,aberbach2022same}. Another potential challenge to our framework is that the definition of telic states as equivalence classes appears inconsistent with reports that human preference in economic choices sometimes exhibit intransitively \cite{tversky1969intransitivity}. However, since decision making relies on sampling from experience distributions, choice anomalies such as preference intransitiviy can be explained in the current framework in terms of random sample deviations (see Supplementary material for details). Ultimately, our framework provides a step towards a formal, epistemological account of how, and why animals represent their environment in particular ways, and not others \cite{song2022minimal}.

\section{Supplementary material}
\label{sec:supp}
\subsection{Learning with telic states}
So far, we have only outlined a framework for deriving telic state representations and illustrated how it can be used to infer goals which best explain the observed behavior. It is natural to ask how can an agent use these goal-derived representations to guide goal-directed behavior. For this, we define a \emph{policy}, $\pi$ as a distribution over actions given the past experience sequence and current observation:
\begin{equation}
\pi(a_i|o_1,a_1,...,o_i).
\end{equation}
Analogously, we define an \emph{environment}, $e$, as a distribution over observations given the past experience sequence: 
\begin{equation}
e(o_i|o_1,a_1,...,a_{i-1}).
\end{equation}
The distribution over experience sequences can be factored, using the chain rule, as follows: 
\begin{equation}
     P_\pi(o_1,a_1,...,o_n,a_n)=P(o_1,a_1,...,o_n,a_n|e,\pi)=\\
    \prod_{i=1}^n e(o_i|o_1,a_1,...,a_{i-1})\pi(a_i|o_1,a_1,...,o_i).
\end{equation} 
Our definition of telic states as goal-induced equivalence classes can now be extended to equivalence between policy-induced experience distributions as follows: 
\begin{equation}
    \pi_1\sim_g\pi_2\iff P_{\pi_1}\sim_gP_{\pi_2}.
\end{equation}
Typically, the environment is assumed to be fixed, and hence not explicitly parameterized in $P_\pi(h)$ above, and the question we are interested in is: how can an agent learn an efficient policy for achieving its goal? In the current framework, we interpret this question as how can an agent increase the likelihood that its policy will generate experiences that belong to a certain telic state $S_i$. In other words, assuming the agent follows a policy $\pi$, and experiences a sample of $N$ sequences, $h_1,...,h_N$, we would like to know how can the agent modify its policy such that the empirical distribution of these samples belongs to a certain telic state $S_i$. Denoting the empirical distribution of sequences by $\hat{P}_\pi$, i.e., $\hat{P}_\pi(h)=\frac{|\{i:h_i=h\}|}{N}$, the answer to this question is given, in the asymptotic limit, by the following result from the theory of large-deviations known as Sanov's theorem:
\begin{equation}
    \lim_{N\to\infty}\frac{1}{N}\log P_\pi^{(N)}(\hat{P}_\pi\in S_i)=-D_{KL}(P^\star||P_\pi),
\end{equation}
where,
\begin{equation}
\label{eq:info-proj}
P_i^\star \equiv arg\min_{P\in S_i} D_{KL}(P||P_\pi),
\end{equation} is the \emph{information projection} of $P_\pi$ onto $S_i$, i.e., the distribution in $S_i$ which is closest (in the KL sense) to $P_\pi$ \cite{cover1999elements}. In other words, the probability that experience sequences sampled from $P_\pi$ will belong to a particular telic state $S_i$ is determined by the KL-divergence between the information projection of $P_\pi$ onto $S_i$ and $P_\pi$ itself. Assuming now the agent follows a policy $\pi_\theta$, parameterized by some $\theta$, the following goal-directed gradient method can be used to update the policy $\pi_\theta$ to one that is more likely to generate experience sequences leading the agent to the telic state $S_i$:
\begin{equation}
    \label{eq:policy_grad_step}
    \theta_{t+1}=\theta_t-\eta\nabla_\theta D_{KL}(P_i^\star||P_{\pi_\theta}).
\end{equation}
\subsubsection{Illustrative example - probability matching in the two-armed bandit}
To illustrate our proposed learning algorithm, we compute the goal-directed policy gradient for a fully-tractable bandit learning problem and show that, in this simple case, minimizing goal-distance yields a commonly reported empirical choice strategy known as probability-matching. We consider a two-armed bandit in which the set of actions is defined as of choosing a left ($L$) or right ($R$) lever and the observations are winning ($1$) or losing ($0$): 
\begin{equation}
    \mathcal{A}=\{L,R\}, \ \mathcal{O}=\{1,0\}.
\end{equation}
The policy, $\pi_\theta$, is parameterized by the probability of choosing action $L$:
\begin{equation}
    \pi_\theta(L)=\theta,\ \pi_\theta(R)=1-\theta.
\end{equation}
The environment $e$ is specified by the probabilities of winning when choosing $L$ or $R$, denoted $p_L$ and $p_R$, respectively:
\begin{equation}
    e(1|L)=p_L, \ e(0|L)=1-p_L; \ e(1|R)=p_R, \ e(0|R)=1-p_R.
\end{equation}
The likelihood that an experience sequence, $h$, will be generated by the policy induced distribution $P_{\pi_\theta}$ can be expressed as:
\begin{equation}
\label{eq:seq_prob}
    P_{\pi_\theta}(h)=\theta^{N^h_L}(1-\theta)^{N^h_R}p_L^{N^h_{L,1}}(1-p_L)^{N^h_L-N^h_{L,1}}p_R^{N^h_{R,1}}(1-p_R)^{N^h_R-N^h_{R,1}},
\end{equation}
where $N^h_L,N^h_R$ are the number of times the agents selected the $L$ and $R$ actions, respectively, and $N^h_{L,1},N^h_{R,1}$ are the number of ``win" observations following $L$ and $R$ choices, respectively. For simplicity, we assume that the agents goal is to maximize the expected number of wins, so that two policies are equivalent if and only if the expected number of wins obtained by following both is equal:
\begin{equation}
\pi_{\theta_1}\sim_g \pi_{\theta_2} \iff \mathbb{E}_{P_{\pi_{\theta_1}(h)}}\left(\sum_i^{N}\mathbf{1}_{h_i=1}\right)= \mathbb{E}_{P_{\pi_{\theta_2}}(h)}\left(\sum_i^{N}\mathbf{1}_{h_i=1}\right),
\end{equation}
where $N=N^h_L+N^h_R$ is the total number of action/observation pairs and $\mathbf{1}_{h_i=1}$ denotes an indicator function which is one if the $i$th observation in $h$ is $1$ and zero otherwise. Thus, for every $j=1,...,N$, the telic state $S_j$ is defined as:
\begin{equation}
    S_j=\{P(h):\text{ s.t. }\mathbb{E}_{P(h)}\left(\sum_i^{N}\mathbf{1}_{h_i=1}\right)=j\},
\end{equation}
and the ``goal-distance" between $P_{\pi_\theta}$ and $S_j$ is given by:
\begin{equation}
D_{KL}(S_j||P_{\pi_\theta})=\min_{P\in S_j} D_{KL}(P||P_{\pi_\theta})=\sum_hP^\star_j(h)\log\frac{P^\star_j(h)}{P_{\pi_\theta}(h)},
\end{equation}
where $P^\star_j(h)$ is the distribution in $S_j$ closest to $P_\pi(h)$ in the KL sense, as defined in Eq.\ref{eq:info-proj} above.
Using Eq.\ref{eq:seq_prob} we can now compute the goal-distance gradient to obtain:
\begin{equation}
    \nabla_\theta D_{KL}(S_i||P_{\pi_\theta})=\frac{\sum_hP^\star_j(h)N_R^h}{1-\theta}-\frac{\sum_hP^\star_j(h)N_L^h}{\theta},
\end{equation}
giving the optimal policy parameter:
\begin{equation}
    \theta^*=\mathbb{E}_{P^\star_j(h)}(\frac{N^h_L}{N^h_L+N^h_R}).
\end{equation}
In words, the policy maximizing the likelihood of reaching telic state $S_i$, is one matching the expected choice probability of $P^{\star}_j(h)$. Interestingly, such ``probability-matching"  strategies have been reported in empirical choice behavior in iterated binary choice tasks \cite{erev2005adaptation}.

\subsection{The flow of experience - transition sensitive goals}
So far, we only considered goals that are sensitive to individual action or observation counts within experience sequences. Indeed, standard reinforcement learning algorithms can be seen as special cases of our framework under a goal of maximizing a (possibly discounted) count of particular observations defined a-priori as rewarding. Real biological agents however, typically pursue more ecological and complex goals, reflecting preferences over higher-order statistics of experience sequences. For example, people engaged in activities such as sports, meditation, artistic creation, game playing or other challenging tasks, often describe the goal of such pursuits as entering a state of ``flow''\cite{csikszentmihalyi1992optimal}, i.e., certain patterns of experience which cannot be reduced to singular actions or outcomes. To explore the potential of our framework, we consider a class of goals defined by maximizing a weighted sum over the number of different possible observation-action and action-observation transitions:
\begin{equation}
\label{eq:transition_count_goal}
    g_{\alpha,\beta}(h)=\sum_{i=1}^{|\mathcal{O}|}\sum_{j=1}^{|\mathcal{A}|}(N^h_{ij}\alpha_{ij}+M^h_{ji}\beta_{ji}),
\end{equation}
where $N^h_{ij}$ denotes the number of transitions between observation $o_t=i$ and action $a_{t+1}=j$, and $M^h_{ji}$ the number of transitions between action $a_{t-1}=j$ and observation $o_t=i$, in a given experience sequence $h=o_1,a_1,...,o_n,a_n$. Under this goal, experience sequences sharing the same empirical transition frequencies will be deemed equivalent by the agent, with the $\alpha_{ij}$ and $\beta_{ji}$ parameters determining the relative weight of different action-observation or observation-action transition, respectively. To make this concrete, we consider a setting where observations depend only on the immediately preceding actions and actions depend only on the immediately preceding observations. The environment can thus be expressed as $e(o_i|o_1,a_1,...,o_{i-1},a_{i-1})=e(o_i|a_{i-1})$, and the policy of the agent as $\pi(a_i|o_1,a_1,...,o_i)=\pi(a_i|o_i)$. The experience distribution induced by the environment $e$ and the policy $\pi$ can be factorized in this case as:
\begin{equation}
    P_\pi(h)=\\
    e(o_1)\pi(a_1|o_1)\prod_{i=2}^ne(o_i|a_{i-1})\pi(a_i|o_i).
\end{equation}
This distribution can be parameterized using the switching probabilities $e_{ji}\equiv e(o_t=i|a_{t-1}=j)$  (with the initial observation distribution, $e_{i}=e(o_1=i)$) and $\pi_{ij}\equiv\pi(a_t=j|o_t=i)$ for $i=1,...,|\mathcal{O}|$ and $j=1,...,|\mathcal{A}|$. Using this notation, we can express the log-probability of experience $h$ given policy $\pi$ and environment $e$ as:
\begin{multline*}
\log{P_\pi(h)} = \log{(e(o_1)\prod_{i=1}^{|\mathcal{O}|}\prod_{j=1}^{|\mathcal{A}|}\pi_{ij}^{N^h_{ij}}e_{ji}^{M^h_{ji}}})=\\
\log{e(o_1)}+\sum_{i=1}^{|\mathcal{O}|}\sum_{j=1}^{|\mathcal{A}|}(N^h_{ij}\log(\pi_{ij})+M^h_{ji}\log(e_{ji})).
\end{multline*}

Assuming sufficiently long sequences, $N^h_{ij}\approx\frac{n}{2}\pi_{ij}$ and $M^h_{ji}\approx \frac{n}{2}e_{ji}$, the goal $g_{\alpha,\beta}(h)$ can be approximated by:
\begin{equation*}
g_{\alpha,\beta}(h)\approx\frac{n}{2}\sum_{i=1}^{|\mathcal{O}|}\sum_{j=1}^{|\mathcal{A}|}(\pi_{ij}\alpha_{ij}+e_{ji}\beta_{ji})=
    -\frac{n}{2}(H(\pi,\pi_{\alpha})+H(e,e_{\beta}))
\end{equation*}
where we define the distributions $\pi_{\alpha}$ and $e_{\beta}$ as:
\begin{equation*}
\pi_{\alpha}(a_t=j|o_{t-1}=i)=\exp(\alpha_{ij}),
\end{equation*}
and
\begin{equation*}
e_{\beta}(o_t=i|a_{t-1}=j)=\exp(\beta_{ji}),
\end{equation*}
with the following normalization constraints on  $\alpha_{ij}$ and $\beta_{ji}$:
\begin{equation*}
\sum_{j=1}^{|\mathcal{A}|}\exp(\alpha_{kj})=\sum_{i=1}^{|\mathcal{O}|}\exp(\beta_{li})=1, \text{ for  $k=1,...,|\mathcal{O}|$ and $l=1,...,|\mathcal{A}|$.}
\end{equation*}
Thus, for a fixed environment, maximizing goal $g_{\alpha,\beta}(h)$ corresponds to minimizing the cross entropy between the true policy $\pi$, and a reference one $\pi_{\alpha}$, induced by the observation-action transition weight parameters $\alpha_{ij}$. This goal therefore can be viewed as inducing a naturally emergent ``information-cost" over policies, extending previous approaches for explaining goal-directed learning in terms of internal complexity costs \cite{amir2020value}. Furthermore, by considering goals that are sensitive to higher-order transition statistics, the current approach can be used to uncover preferences over certain flows of experience (action-observation subsequences), beyond mere accumulation of externally-determined ``rewarding" outcomes.


\begin{thebibliography}{23}
\providecommand{\natexlab}[1]{#1}
\providecommand{\url}[1]{\texttt{#1}}
\expandafter\ifx\csname urlstyle\endcsname\relax
  \providecommand{\doi}[1]{doi: #1}\else
  \providecommand{\doi}{doi: \begingroup \urlstyle{rm}\Url}\fi

\bibitem[Aberbach-Goodman et~al.(2022)Aberbach-Goodman, Buaron, Mudrik, and Mukamel]{aberbach2022same}
S.~Aberbach-Goodman, B.~Buaron, L.~Mudrik, and R.~Mukamel.
\newblock Same action, different meaning: neural substrates of action semantic meaning.
\newblock \emph{Cerebral Cortex}, 32\penalty0 (19):\penalty0 4293--4303, 2022.

\bibitem[Amir et~al.(2020)Amir, Suliman-Lavie, Tal, Shifman, Tishby, and Nelken]{amir2020value}
N.~Amir, R.~Suliman-Lavie, M.~Tal, S.~Shifman, N.~Tishby, and I.~Nelken.
\newblock Value-complexity tradeoff explains mouse navigational learning.
\newblock \emph{PLOS Computational Biology}, 16\penalty0 (12):\penalty0 e1008497, 2020.

\bibitem[Balcetis and Dunning(2006)]{balcetis2006see}
E.~Balcetis and D.~Dunning.
\newblock See what you want to see: motivational influences on visual perception.
\newblock \emph{Journal of personality and social psychology}, 91\penalty0 (4):\penalty0 612, 2006.

\bibitem[Bowling et~al.(2022)Bowling, Martin, Abel, and Dabney]{bowling2022settling}
M.~Bowling, J.~D. Martin, D.~Abel, and W.~Dabney.
\newblock Settling the reward hypothesis.
\newblock \emph{arXiv preprint arXiv:2212.10420}, 2022.

\bibitem[Cover(1999)]{cover1999elements}
T.~M. Cover.
\newblock \emph{Elements of information theory}.
\newblock John Wiley \& Sons, 1999.

\bibitem[Csikszentmihalyi and Csikszentmihalyi(1992)]{csikszentmihalyi1992optimal}
M.~Csikszentmihalyi and I.~S. Csikszentmihalyi.
\newblock \emph{Optimal experience: Psychological studies of flow in consciousness}.
\newblock Cambridge university press, 1992.

\bibitem[Dunne(2004)]{dunne2004foundations}
J.~D. Dunne.
\newblock \emph{Foundations of Dharmakirti's philosophy}.
\newblock Simon and Schuster, 2004.

\bibitem[Erev and Barron(2005)]{erev2005adaptation}
I.~Erev and G.~Barron.
\newblock On adaptation, maximization, and reinforcement learning among cognitive strategies.
\newblock \emph{Psychological review}, 112\penalty0 (4):\penalty0 912, 2005.

\bibitem[Fogassi et~al.(2005)Fogassi, Ferrari, Gesierich, Rozzi, Chersi, and Rizzolatti]{fogassi2005parietal}
L.~Fogassi, P.~F. Ferrari, B.~Gesierich, S.~Rozzi, F.~Chersi, and G.~Rizzolatti.
\newblock Parietal lobe: from action organization to intention understanding.
\newblock \emph{Science}, 308\penalty0 (5722):\penalty0 662--667, 2005.

\bibitem[Hume(1896)]{hume1896treatise}
D.~Hume.
\newblock \emph{A treatise of human nature}.
\newblock Clarendon Press, 1896.

\bibitem[Kaelbling(1993)]{kaelbling1993learning}
L.~P. Kaelbling.
\newblock Learning to achieve goals.
\newblock In \emph{IJCAI}, volume~2, pages 1094--8. Citeseer, 1993.

\bibitem[Langdon et~al.(2019)Langdon, Song, and Niv]{langdon2019uncovering}
A.~J. Langdon, M.~Song, and Y.~Niv.
\newblock Uncovering the ‘state’: Tracing the hidden state representations that structure learning and decision-making.
\newblock \emph{Behavioural Processes}, 167:\penalty0 103891, 2019.

\bibitem[Leong et~al.(2019)Leong, Hughes, Wang, and Zaki]{leong2019neurocomputational}
Y.~C. Leong, B.~L. Hughes, Y.~Wang, and J.~Zaki.
\newblock Neurocomputational mechanisms underlying motivated seeing.
\newblock \emph{Nature human behaviour}, 3\penalty0 (9):\penalty0 962--973, 2019.

\bibitem[Li et~al.(2006)Li, Walsh, and Littman]{li2006towards}
L.~Li, T.~J. Walsh, and M.~L. Littman.
\newblock Towards a unified theory of state abstraction for {MDP}s.
\newblock In \emph{AI\&M}, 2006.

\bibitem[Minsky(1967)]{minsky1967computation}
M.~Minsky.
\newblock Computation: Finite and infinite machines prentice hall.
\newblock \emph{Inc., Engelwood Cliffs, NJ}, 1967.

\bibitem[Molinaro and Collins(2023)]{molinaro2023goal}
G.~Molinaro and A.~G.~E. Collins.
\newblock A goal-centric outlook on learning.
\newblock \emph{Trends in Cognitive Sciences}, 2023.

\bibitem[Niv(2019)]{niv2019learning}
Y.~Niv.
\newblock Learning task-state representations.
\newblock \emph{Nature Neuroscience}, 22\penalty0 (10):\penalty0 1544--1553, 2019.

\bibitem[Roesch et~al.(2006)Roesch, Taylor, and Schoenbaum]{roesch2006encoding}
M.~R. Roesch, A.~R. Taylor, and G.~Schoenbaum.
\newblock Encoding of time-discounted rewards in orbitofrontal cortex is independent of value representation.
\newblock \emph{Neuron}, 51\penalty0 (4):\penalty0 509--520, 2006.

\bibitem[Rosch and Mervis(1975)]{rosch1975family}
E.~Rosch and C.~B. Mervis.
\newblock Family resemblances: Studies in the internal structure of categories.
\newblock \emph{Cognitive Psychology}, 7\penalty0 (4):\penalty0 573--605, 1975.

\bibitem[Siderits et~al.(2011)Siderits, Tillemans, and Chakrabarti]{siderits2011apoha}
M.~Siderits, T.~J. Tillemans, and A.~Chakrabarti.
\newblock \emph{Apoha: Buddhist nominalism and human cognition}.
\newblock Columbia University Press, 2011.

\bibitem[Song et~al.(2022)Song, Takahashi, Burton, Roesch, Schoenbaum, Niv, and Langdon]{song2022minimal}
M.~Song, Y.~K. Takahashi, A.~C. Burton, M.~R. Roesch, G.~Schoenbaum, Y.~Niv, and A.~J. Langdon.
\newblock Minimal cross-trial generalization in learning the representation of an odor-guided choice task.
\newblock \emph{PLoS Computational Biology}, 18\penalty0 (3):\penalty0 e1009897, 2022.

\bibitem[Tversky(1969)]{tversky1969intransitivity}
A.~Tversky.
\newblock Intransitivity of preferences.
\newblock \emph{Psychological review}, 76\penalty0 (1):\penalty0 31, 1969.

\bibitem[Wittgenstein(1953)]{wittgenstein1953philosophical}
L.~Wittgenstein.
\newblock \emph{Philosophical investigations}.
\newblock Wiley-Blackwell, New York, NY, USA, 1953.

\end{thebibliography}






\end{document}